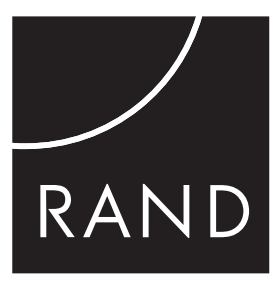

Expert Insights

PERSPECTIVE ON A TIMELY POLICY ISSUE

PATRICIA PASKOV, CHRISTOPHER RODRIGUEZ, SUNISHCHAL DEV, STEPHEN CASPER

# Open-Weight AI Models Require Proportional Evaluation Approaches

*This publication has completed RAND's research quality-assurance process but was not professionally copyedited.*

# About This Paper

Open-weight AI models (OWMs) – models released with publicly-available weights – are distributing rapidly and approaching the performance levels of leading closed-weight AI models (CWMs). While OWMs offer substantial scientific and economic benefits, their release introduces distinct risk factors for which existing evaluation practices, largely designed for CWM deployment, fail to account. In this paper, we argue that these risk factors demand distinct proportional evaluation (PE) approaches: evaluating without system-level safeguards (PE1), assessing robustness to modifications that undo model-level safeguards (PE2), testing selective capability amplification (PE3), and proxying worst-case misuse (PE4). We systematically review current evaluation practices of OWMs released in 2025 through April 2026, finding that only one of the 37 families of models reviewed fulfills PE1-4 and most do not fulfill any. This paper targets policymakers, funders, and researchers involved in AI evaluation. As OWMs grow increasingly capable, their evaluation warrants close attention from developers, funders, and governance bodies alike.

## Center on AI, Security, and Technology

RAND Global and Emerging Risks is a division of RAND that delivers rigorous and objective public policy research on the most consequential challenges to civilization and global security. This work was undertaken by the division's Center on AI, Security, and Technology, which aims to examine the opportunities and risks of rapid technological change, focusing on artificial intelligence, security, and biotechnology. For more information, contact cast@rand.org.

## Funding

This effort was independently initiated and conducted within the Center on AI, Security, and Technology using income from operations and gifts and grants from philanthropic supporters. A complete list of donors and funders is available at www.rand.org/CAST. RAND clients, donors, and grantors have no influence over research findings or recommendations.

## Acknowledgments

We thank Stella Biderman, Ben Bucknall, Charles Foster, Avijit Ghosh, Jeffrey Lee, Kellin Pelrine, Aviya Skowron, Cozmin Ududec, and Sophie Williams for feedback; Lee Nilsson and Brittany Thomas for support; and Steph Guerra, Sella Nevo, Henry Willis, and Bill Zito for their leadership.

# Contents

# Tables

# Chapter 1. Introduction

Open-weight AI models (OWMs) – models released with publicly available weights – are being developed and deployed at an accelerating pace (Bhandari et al., 2025). These systems offer to expand and democratize open science by enabling widespread collaboration, rapid experimentation, and community-driven innovation (Bommasani et al., 2024; Eiras et al., 2024; Francois et al., 2025; Jernite et al., 2025; Kapoor et al., 2024; Miller et al., 2025; NTIA, 2024; Seger and O'Dell, 2024; UK AISI, 2025c). Yet the same properties that make OWMs valuable also introduce distinct risk factors (RF), presenting security and governance challenges unique from those posed by closed-weight AI models (CWMs). In particular, system-level safeguards are removable (RF1), model-level safeguards are modifiable (RF2), dangerous capabilities amplification faces fewer post-release restrictions (RF3), and model weights can spread easily and reversibly (RF4). In this paper, we examine how these unique risk factors demand proportional evaluation (PE) approaches for OWMs across domains.

Recent research suggests that the performance of OWMs across most common metrics and benchmarks lags 6-12 months behind those of CWMs (Cottier, Martemianova, and Owen, 2024; Maslej et al., 2025), with some estimates as low as four months (UK AISI, 2025c). Pre-deployment evaluations indicate that leading CWMs are approaching capability thresholds associated with serious safety risks (Anthropic, 2025a; Google DeepMind, 2025; OpenAI, 2025a),[1] and security incident reports show attempts by a range of threat actors to misuse CWMs for cyber and influence operations (Anthropic, 2025b; Anthropic, 2025c; OpenAI, 2025b). Given the narrow performance gap, OWMs may soon attain comparable levels of dangerous capabilities. What's more, OWMs can be run locally, without any ability for developers or third parties to observe how they are being used (Vicens, 2026).

Across the world, AI governance frameworks increasingly rely on evaluations of AI to guide risk assessment, mitigation, and deployment decisions (Appendix B; UKDSIT, 2024; Frontier Model Forum, 2024, 2025; European Commission, 2025; Senate Bill-53; RAISE Act, 2025); and prior work highlights evaluations as a key tool for characterizing risks of and informing risk mitigation strategies for OWMs (Srikumar et al., 2024; Casper et al., 2025). Despite this, OWMs are often released with little to no information about whether or how they were evaluated (Casper et al., 2025; Wan et al., 2025). When OWM evaluations are reported, they often focus on safety-trained, production-ready variants, overlooking additional risk factors introduced by OWM releases (e.g. Yang et al., 2025; Meta, 2025a).

---

[1] OpenAI's GPT-5-thinking, for example, was treated as High Capability in cyber/chemical domains, Claude Opus 4 was deployed under AI Safety Level 3, and Gemini 2.5 Deep-Think has enough technical knowledge in certain CBRN scenarios and stages to be considered at early alert threshold.

In this paper, we examine how the distinct risk factors of OWMs call for proportional evaluation approaches. Chapter 2 highlights four risk factors (RF) that demand proportional evaluation (PE) approaches to achieve adequate OWM evaluation (Table 1.1), drawing from a review of OWM literature and model and system cards and our team's expertise in machine learning, engineering, and AI evaluation. Chapter 3 reviews current evaluation practices of OWMs released between January 2025 and February 2026 using greater than 10^24 FLOPs. Chapter 4 concludes. By highlighting the distinct risk factors of OWMs and outlining corresponding PE approaches, we seek to provide a foundation on which comprehensive best practices for OWM evaluations can emerge.

**Table 1.1. Open-weight Model Risk Factors and Corresponding Proportional Evaluation Approaches**

| | Open-Weight Model Risk Factor (RF) | Proportional Evaluation (PE) Approach |
|---|---|---|
| Section 2.1 | RF1: System-level safeguards are removable or non-existent | PE1: Evaluate without system-level safeguards |
| Section 2.2 | RF2: Model-level safeguards are modifiable | PE2: Assess robustness to modifications designed to undo model-level safeguards |
| Section 2.3 | RF3: Dangerous capabilities amplification faces fewer post-release restrictions | PE3: Assess selective capability amplification via fine-tuning and tool use |
| Section 2.4 | RF4: (cross-cutting RF1-3): Model weights can spread easily and irreversibly | PE4: (cross-cutting PE1-3): Proxy worst-case feasible misuse |

# Chapter 2. Evaluating Open-weight Models: Risk Factors and Proportional Approaches

Until DeepSeek-R1's release in January 2025, highly capable AI systems consisted of a set of CWMs from a small number of developers. Accordingly, evaluations and risk management agendas during this period largely came to reflect this frontier CWM regime. Many dangerous-capability evaluations – and the reporting standards that structure them, such as system and model cards – emerged primarily with CWMs in mind (e.g. Shevlane et al., 2023). As such, approaches for CWM safety rely on assumptions that do not hold for OWMs, for which system-level controls are removable or non-existent (RF1), model-level safeguards can be modified (RF2), dangerous capability amplification faces fewer post-release restrictions (RF3), and model weights can spread easily and irreversibly (RF4). Evaluations designed for CWMs may therefore systematically understate open-weight risks.

These structural gaps motivate a set of proportional evaluation approaches, calibrated to the marginal risk a model plausibly introduces (Mougan et al., 2026). Proportionality cuts both ways: many OWMs lag leading CWMs in capabilities and may not, on their own, warrant exhaustive stress-testing. Holding capability constant, however, we argue that RF1–4 each introduce distinct evaluation gaps that PE1–4 are designed to address. The degree of implementation of PE1-4 depends on the degree to which a given OWM plausibly expands marginal risk.[2]

The approaches we identify in this section apply across evaluation paradigms, including multiple-choice question-answer benchmarks, human uplift studies, long-form task-based agentic evaluations, and adversarial stress-testing. We do not revisit evaluation practices that apply equally to CWMs and OWMs, instead focusing exclusively on additional considerations for OWMs.

## 2.1 System-level Safeguards

### *Risk Factor 1 (RF1): System-level safeguards are removable or non-existent*

Many AI systems rely on system-level safeguards external to the central model. These include system prompts and wrappers that promote safe behavior (e.g. Rebedea et al., 2023), content filters that block harmful inputs or outputs (Ahmed and Iyer, 2025; AWS Bedrock 2026; Inan et al., 2023; Meta, 2025b; McKenzie et al., 2026; Microsoft Azure 2026; Sharma et al., 2025), behavioral probes that monitor for malicious activity (Kramar et al., 2026; MacDiarmid et al., 2024), and moderation models that edit or suppress suspicious outputs (Greenblatt et al., 2023).

[2] Recent policy approaches illustrate both the appeal and the fragility of proportional reasoning: for example, Commitment 6 of the EU AI Office Code of Practice exempts models from certain security obligations if they fall below at least one publicly downloadable OWM, effectively making OWMs a moving baseline for downstream requirements (European Commission, 2025). While pragmatic in intent, such competitor-relative thresholds risk gradual standards erosion if reference models are weakly evaluated or uncertainty is ignored, a dynamic akin to the "boiling frog" problem described by Bateman et al. (2024) and highlighted in Williams et al. (2025).

In CWMs, system-level safeguards can be enforced, monitored, updated, and iteratively refined server-side within a centralized deployment stack (e.g. as in Google, 2025; OpenAI, 2024). By contrast, OWM safeguards can be trivially removed by users (Casper et al., 2025) and resulting systems can be hosted and distributed. Accordingly, behavior observed during pre-deployment testing under system-level safeguards may not reflect how OWMs behave in real-world use.

### *Proportional Evaluation Approach 1 (PE1): Evaluate without system-level safeguards*

Evaluating OWMs with system-level safeguards may lead to inaccurate capability and risk estimates. For example, while AI image and video generators are often released alongside safety filters, these can be and frequently are disabled by users wishing to produce restricted content (Kamachee et al., 2025; Rando et al., 2022). As a result, OWM evaluations that test minimally wrapped models – void of system-level prompts or safety wrappers, content filters, behavioral probes, or moderation models – may better characterize the capabilities and risks of OWMs independent of deployment-layer safeguards.

## 2.2 Model-level Safeguards

### *Risk Factor 2 (RF2): Model-level safeguards are modifiable*

Beyond application-level controls, many AI systems rely on model-level safeguards baked into the model's parameters. These safeguards are typically introduced via techniques such as supervised fine-tuning (e.g. Guan et al., 2025; Wallace et al., 2024), reinforcement learning from human feedback (RLHF) (Bai et al., 2022; Christiano et al., 2017; Guan et al., 2024; Ouyang et al., 2022), and adversarial training (e.g. Altinisik et al., 2023; Hubinger et al., 2024).

In CWMs, end users cannot directly access or modify model parameters or internal states. OWM release, by contrast, grants users direct access to model weights and activations, enabling users to weaken, bypass, or remove safeguards after release (AI Security Institute, 2026; Casper et al., 2025). In particular, safeguards may be weakened through two key mechanisms: *weight-level modifications*[3] and *activation-level modifications* at inference time.[4]

### *Proportional Evaluation Approach 2 (PE2): Assess robustness to modifications designed to undo model-level safeguards*

Assessing OWM behavior with model-level safeguards may not accurately predict performance in the wild. Model-level safeguard removal from OWMs is both prevalent and non-trivial: for example, thousands of OWMs have been fine-tuned to disable safeguards (Casper et al., 2025). At the time of this paper's writing, a popular platform hosts over 8000 "uncensored" or "abliterated" models, or models with removed refusal-based safeguards (Hugging Face, 2026a, 2026b).[5] OWMs without model-level safeguards have been shown to comply with unsafe prompts at far higher rates than unmodified models across multiple harm domains (Dombrowski et al., 2025; Sokhansanj, 2025).[6]

To assess real-world risk, OWM evaluations can stress-test model-level safeguards through tampering evaluations (Casper et al., 2024, 2025; Che et al., 2025; Gal, 2024; Gal and Casper, 2025; Hossain et al., 2026; O'Brien et al., 2025), which assess how model behavior changes under realistic modifications to safeguards. In particular, these evaluations may assess the model under:

- *weight-level modifications,* such as malicious fine-tuning (e.g. Lermen et al., 2023; Li et al., 2024, Murphy et al., 2025; Zhan et al., 2024), non-malicious fine-tuning (e.g. Betley et al., 2025; Che et al., 2025; He et al., 2024; Pandey et al., 2025; Qi et al., 2023), and training-free weight edits (e.g., Young, 2025) and
- *activation-level modifications* at inference time that bypass refusals or extract suppressed information (e.g. Arditi et al., 2024; Lermen et al., 2024, Panickssery et al., 2023; Seyitoğlu et al., 2024; Struppek et al., 2026).

---

[3] Weight-level modifications may include 1) malicious fine-tuning, which deliberately targets previously learned safety behaviors (Lermen et al., 2023; Li et al., 2024; Qi et al., 2023; Zhan et al., 2024), 2) fine-tuning on non-malicious data which has also been shown to degrade safeguards (Betley et al., 2025; Che et al., 2025; He, Xia, and Henderson, 2024; Pandey et al., 2026; Qi et al., 2023), and 3) training-free interventions such as "refusal ablation," which remove refusal behavior through direct parameter modification without gradient descent (Young, 2026).

[4] Under these modifications, internal activations are manipulated without model weight modification. These methods can extract "unlearned" dangerous information (Seyitoğlu et al., 2024) or override mechanisms on a per-query basis (Arditi et al., 2024; Lermen, Dziemian, and Pimpale, 2024; Panickssery et al., 2024).

[5] Other providers host these models that can be used in a chatbot style interface without any need for the user to provision GPU machines to host the model themselves (Featherless, undated).

[6] Dombrowski et al. (2025) document a growing "safety gap" between baseline and safeguard-removed versions of frontier OWMs (e.g., Llama-3, Qwen-2.5) in biological, chemical, and cyber domains, with the gap increasing as model scale grows, indicating that internal safeguards to date become less effective precisely as underlying capabilities increase. Sokhansanj (2025) finds that models with safety removal or "uncensoring" comply with unsafe prompts at substantially higher rates than their unmodified counterparts across domains including fraud, psychological manipulation, physical harm, and social disorder.

## 2.3 Capability Amplification

### *Risk Factor 3 (RF3): Dangerous capabilities amplification faces fewer post-release restrictions*

Developers of AI systems can limit dangerous capabilities through pretraining data curation (Liu et al. 2024; O'Brien et al., 2025) and targeted unlearning during post-training. While CWMs also support fine-tuning and tool use, providers can impose restrictions on data, objectives, and tools (Chen and Yang, 2023; Eldan and Russinovich, 2024; Liu et al., 2025; Yao, Xu, and Liu, 2023; e.g. Microsoft Azure, 2026). OWMs face no such constraints: users may fine-tune OWMs for specific high-risk domains, reintroduce harmful data, or integrate models into tool-using or agentic systems without oversight (AI Security Institute, 2026; Casper et al., 2025).[7] These interventions can selectively amplify dangerous capabilities beyond those observed at release, potentially exceeding what is achievable in CWM settings (Che et al., 2025; Hossain et al., 2026; Pandey et al., 2026).

### *Proportional Evaluation Approach 3 (PE3): Assess selective capability amplification via fine-tuning and tool use*

OWM capabilities measured at release, absent fine-tuning, may substantially underestimate what can be achieved downstream. For example, researchers used reinforcement learning to fine-tune Llama-3.1-8B on cryptographic challenges and observed improved performance on unseen cybersecurity tasks (Muzsai et al., 2025). Because OWM capability profiles can change substantially post-release, evaluations that assess only baseline performance are particularly prone to fail in capturing the upper bound of risk associated with selective capability amplification.[8]

Accordingly, evaluations that test models under attempts to amplify their capabilities in harmful domains are likely to provide more informative evidence about real-world risk (Wallace et al., 2025). Where relevant, this includes fine-tuning on high-risk domain data (e.g., cyber offense or biosecurity-relevant tasks, as in Wallace et al., 2025); tool use, such as web search and code execution (O'Brien et al., 2025); and test-time scaling via extended reasoning budgets (Liu et al., 2025; Snell et al., 2024).

---

[7] We note, however, that jailbreaks have been shown to circumvent these restrictions in some cases (e.g. Bowen et al., 2025; Murphy et al., 2025).

[8] While this challenge exists for CWMs using black-box elicitation, open-weight releases enable more extensive post-deployment modification, compounding the difficulty of worst-case assessment.

## 2.4 Irreversibility

### *RF4 (cross-cutting RF1-3): Model weights can spread easily and irreversibly*

CWMs are typically accessed through centralized hosting or APIs, which allow providers to revoke access, roll back deployments, patch vulnerabilities, or impose additional controls in response to emerging risks. The private nature of closed model weights makes unauthorized access and replication much more difficult (Nevo et al., 2024).

In contrast, once OWMs are publicly released, control over their distribution is lost. Even if hosting platforms remove files or impose restrictions (e.g. as in Kilcher, 2023), copies can persist and continue to circulate offline or across alternative platforms (AI Security Institute, 2026; Casper et al., 2025; Maiberg, 2025).[9] Centralized monitoring or moderation is infeasible for open-weight systems (AI Security Institute, 2026; Casper et al., 2025; Dombrowski et al., 2025).

### *PE4 (cross-cutting PE1-3): Proxy worst-case feasible misuse*

Because OWM release is generally irreversible, pre-deployment evaluation carries greater decision weight than for CWMs: vulnerabilities identified after release cannot be patched and access cannot be revoked. As irreversibility exacerbates the consequences of RF1–3, well-resourced and technically skilled actors capable of removing safeguards and enhancing dangerous capabilities warrant heightened consideration in OWM scenarios. With reasonable proportionality, evaluations would ideally replicate or simulate what those actors could plausibly do. Actor resources are not unbounded, however: beyond a certain threshold, a sufficiently resourced actor could simply develop a capable model from scratch, providing a natural upper bound on the evaluation regime.

OWM evaluations may consider distinct threat actor profiles ranging from hobbyist to state-backed actors and conduct evaluations with corresponding resources.[10] For example, OpenAI used a 7-figure USD budget in GPU hours for malicious finetuning exercises for pre-deployment evaluations of GPT-OSS (Wallace et al., 2025). Such evaluations may be expensive and operationally demanding, especially for smaller or nonprofit OWM developers. We therefore emphasize PE4 as a proportional approach that scales with plausible marginal risk and whose implementation may rely on shared infrastructure. Further, consistent with Appendix 3.2 of the

---

[9] Additionally, if scaffolded as agents with effectively computer- and internet-use-complete action spaces, open-weight models could propagate themselves over unsecured networks (Black et al., 2024; Pan et al., 2025). If small-but-capable OWMs start to replicate themselves indefinitely, this could lead to smart malware that could be quite difficult to remove (UK AISI, 2026).

[10] For evaluations emulating state actors, resources and budget may be large, but smaller training runs can be conducted and improvements extrapolated. For example, following malicious fine-tuning (MFT) and model-tampering work, developers can train lightweight anti-refusal and domain-specialist fine-tunes using tens to hundreds of high-quality demonstrations, and where appropriate, allocate larger reinforcement-learning budgets, for example, on the order of thousands of GPU-hours, in realistic tool-using environments to approximate well-resourced state-backed actors.

Code of Practice (European Commission 2025), PE4 can be informed by the conditions of identified, domain-specific threat scenarios with a defined threat actor, misuse objective, permitted modifications and tools, resource assumptions, and success criteria. Because the threat surface is broad and first-party evaluators face inherent incentive pressures that can narrow the scope of internal testing, third-party evaluators may be particularly well-positioned to support evaluations of worst-case behaviors (Brundage et al., 2026; UK AISI, 2025b; Wallace et al., 2025).

# Chapter 3. Current Practices

We review current evaluation practices of OWMs released between January 2025 and April 2026 using greater than 10^24 FLOPs, drawing from the Epoch (2026) database. This threshold captures high-compute models that exceed the EU AI Act's ≥10^23 FLOP threshold for general-purpose AI models while remaining just below the ≥10^25 FLOP level associated with systemic-risk designations, allowing us to analyze models approaching emerging regulatory compute thresholds.[11] We first extract publicly available system cards, model cards, and technical reports for each model. We then assess whether developers conduct dangerous capability evaluations and whether they report implementing the PE approaches discussed in Chapter 2. For detailed coding rules and model-by-model results, refer to Appendix A.

**Table 3.1. Adoption of proportional evaluation (PE) approaches among open weight models and families released between January 2025 and April 2026.**

| Metric | Models (n=52) | Families (n=37) |
|---|---|---|
| Reports any safety evaluation | 24 (46%) | 18 (49%) |
| Reports CBRN/cyber safety evaluation | 6 (12%) | 5 (14%) |
| PE1: Evaluates without system-level safeguards | 18 (35%) | 14 (38%) |
| PE2: Assesses robustness to modifications designed to undo model-level safeguards | 4 (8%) | 4 (11%) |
| PE3: Assess selective capability amplification via fine-tuning and tool use | 1 (2%) | 1 (3%) |
| PE4: Assumes irreversible release and worst-case feasible misuse. | 1 (2%) | 1 (3%) |

Out of 37 OWM families trained using greater than 10^24 FLOPs released between January 2025 and April 2026, 18 (49%) report conducting any safety evaluations,[12] while only 5 (14%) report conducting dangerous capability evaluations in chemical-biological-radiological-nuclear (CBRN) or cyber domains. Many model families (38%) report evaluations without system-level safeguards (PE1), though this is largely because safeguards are not reported to be bundled into their systems in the first place. Very few model families (11%) report evaluations with modifications designed to undo model-level safeguards (PE2). Only one model family (3%), OpenAI's GPT-OSS, reports evaluations with amplified dangerous capabilities (PE3) and evaluations that proxy worst-case feasible misuse (PE4).

[11] Under the EU AI Act, models trained with ≥10^23 FLOPs are generally treated as general-purpose AI models, triggering baseline transparency and documentation obligations for providers placing them on the EU market. More stringent evaluation-related requirements arise at higher compute thresholds, including 10^25 FLOPs for the AI Act's systemic-risk designation and $10^{26}$ FLOPs under the proposed U.S. RAISE Act. As of 2025, only one released open-weight model, Llama 4 Behemoth (preview), is estimated to have crossed the 10^25 FLOP threshold, though continued scaling suggests additional models may soon reach these levels.

[12] E.g. often in the form of simply running on safety benchmarks like TruthfulQA or HarmBench (Lin, Hilton, and Owain, 2021; Mazeika et al., 2024).

# Chapter 4. Conclusion

OWMs present both promising opportunities and significant risks. Evaluations empower decision-makers to measure these risks and prevent harms from OWMs (Casper et al., 2025; Srikumar et al., 2024). Yet the risk factors that distinguish OWMs from CWMs – removable system-level safeguards (RF1), modifiable model-level safeguards (RF2), post-release capability amplification (RF3) and easy and irreversible model weight spread (RF4) – demand evaluation approaches beyond those designed for CWMs.

In this paper, we identified four proportional evaluation (PE) approaches that correspond to these risk factors: evaluating models without system-level safeguards (PE1), assessing robustness to modifications designed to undo model-level safeguards (PE2), testing selective capability amplification (PE3), and proxying worst-case misuse (PE4). Our systematic review of OWM releases in 2025 through April 2026 reveals substantial gaps between these approaches and current practice: while 49% of model families report safety evaluations, only 14% report dangerous capabilities in CBRN or cyber domains. Further, 38% report evaluating models without system-level safeguards (PE1) and only 11% report evaluations with modifications to model-level safeguards (PE2). Only one model family (3%), OpenAI's GPT-OSS, reports evaluations with with amplified dangerous capabilities (PE3) and evaluations that proxy worst-case feasible misuse (PE4).

Beyond PE1–4, we highlight transparent documentation as essential to both the implementation and assessment of these approaches. Whether an evaluation meaningfully satisfies PE1–4 will often depend on how the evaluation is designed and implemented and the research question it seeks to answer, all of which can be communicated by thorough transparent documentation.[13] This is especially true for PE3 and PE4, where poorly specified or bad-faith implementation can create a misleading appearance of rigor. To this end, independent verification can further strengthen the implementation and assessment of PE1–4.

The same properties that make OWMs valuable – publicly accessible weights enabling widespread collaboration, rapid experimentation, and community-driven innovation – are precisely those that introduce RF1–4. The benefits of open-weight release extend even to organizations that do not themselves release open models: CWM developers also rely on OWMs for reproducible safety research, underscoring the scientific value of maintaining open alternatives across capability levels (Anthropic, 2026). The risks and benefits of OWM release are therefore not separable. Evaluation frameworks that account for RF1–4 can support realizing the benefits of openness responsibly.

[13] Detailed reporting of evaluation methods and results may itself raise information-hazard concerns. Evaluation teams should therefore balance transparency with safety, making use of secure information transfer as appropriate. Frontier Model Forum, for example, proposes a three-tiered approach to information-sharing on AI-biology evaluations, where only trusted, expert networks receive access to the most sensitive information (Frontier Model Forum 2025).

As OWMs continue to approach dangerous capability thresholds, aligning evaluation practice with the distinct risks of OWM release remains critical to informed governance. Similar considerations may extend to other emerging open-weight domains, such as AI-enabled biotechnology tools, where safeguards and evaluation practices remain comparatively underdeveloped (Epoch AI, 2025).

# Appendix A. Systematic Review Results and Methodology

## Table A.1. Assessment of Model Safety Evaluations

| Family | Organization | # Models | Safety | CBRN/Cyber | SOTA1 | SOTA2 | SOTA3 | SOTA4 |
|---|---|---|---|---|---|---|---|---|
| Qwen2.5 | Alibaba | 1 | N | N | N | N | N | N |
| Qwen3 | Alibaba | 7 | N | N | N | N | N | N |
| Qwen3.5 | Alibaba | 1 | N | N | N | N | N | N |
| OLMo 2 | Allen Institute for AI | 1 | Y | N | Y | N | N | N |
| Ling | Ant Group | 2 | Y | N | Y | N | N | N |
| Trinity | Arcee | 1 | N | N | N | N | N | N |
| Emu3.5 | BAAI | 1 | N | N | N | N | N | N |
| ERNIE-4.5-A | Baidu | 1 | N | N | N | N | N | N |
| Seed-OSS | ByteDance | 1 | Y | N | Y | N | N | N |
| Foundation-sec-8b | Cisco | 1 | N | N | N | N | N | N |
| DeepSeek-R1 | DeepSeek | 2 | Y | N | Y | N | N | N |
| DeepSeek-V3.1 | DeepSeek | 2 | N | N | N | N | N | N |
| DeepSeek-V3.2 | DeepSeek | 1 | N | N | N | N | N | N |
| Apertus | ETH Zurich,EPFL,CSCS,Swisscom | 1 | Y | Y | Y | N | N | N |
| Gemma 4 | Google DeepMind | 1 | Y | N | Y | Y | N | N |
| Pangu | Huawei | 1 | Y | N | N | N | N | N |
| Granite-4.0-H | IBM | 1 | Y | N | Y | N | N | N |
| EXAONE 4.0 | LG AI Research | 1 | N | N | N | N | N | N |
| K-EXAONE | LG AI Research | 1 | Y | Y | Y | Y | N | N |
| LongCat | Meituan Inc | 1 | Y | N | N | N | N | N |
| Llama 4 | Meta AI | 2 | Y | Y | N | N | N | N |
| MiniMax-M1 | MiniMax | 2 | N | N | N | N | N | N |
| Mistral | Mistral AI | 1 | N | N | N | N | N | N |
| Kimi K2 | Moonshot | 1 | Y | Y | Y | Y | N | N |
| Kimi K2.5 | Moonshot | 1 | N | N | N | N | N | N |
| Llama Nemotron | NVIDIA | 1 | N | N | N | N | N | N |
| Nemotron-H | NVIDIA | 1 | Y | N | Y | N | N | N |
| NVIDIA-Nemotron-Nano-v2 | NVIDIA | 2 | Y | N | N | N | N | N |
| gpt-oss | OpenAI | 1 | Y | Y | Y | Y | Y | Y |
| dots.llm1 | Rednote | 1 | N | N | N | N | N | N |
| Falcon-H1 | Technology Innovation Institute | 1 | N | N | N | N | N | N |
| MiMo | Xiaomi Corp | 2 | N | N | N | N | N | N |
| YandexGPT 5 | Yandex | 1 | N | N | N | N | N | N |
| GLM-4 | Z.ai | 1 | Y | N | Y | N | N | N |
| GLM-4.5 | Z.ai | 3 | Y | N | Y | N | N | N |
| GLM-4.6 | Z.ai | 1 | Y | N | Y | N | N | N |
| GLM 5 | Z.ai | 1 | N | N | N | N | N | N |

We use the Epoch AI Model Database (Epoch, 2026) to identify open-weight models released post 2025 with training compute at or exceeding 10^24 FLOPs. We included Gemma 4 even though it technically did not have an official estimate of training compute in that database as of release of this piece due to the likelihood that it meets our threshold. We use available documentation for each model (usually from the technical report or HuggingFace model card) to assess whether they perform any type of safety evaluation, CBRN evaluations specifically, and each SOTA evaluation detailed in this report (testing without system-level safeguards, testing without model-level safeguards, attempting to amplify capabilities, and assuming irreversible release). We only have access to public information, so some labs may have performed additional safety training that they chose not to disclose.

# Appendix B. Excerpts from Relevant Standards and Regulations

## The EU AI Act Codes of Practice

The following are excerpts from the Safety and Security Chapter (European Commission, 2025) of the EU AI Act General Purpose AI Code of Practice.

### *Measure 3.2 Model evaluations*

Signatories will conduct at least state-of-the-art model evaluations in the modalities relevant to the systemic risk to assess the model's capabilities, propensities, affordances, and/or effects, as specified in Appendix 3.

Signatories will ensure that such model evaluations are designed and conducted using methods that are appropriate for the model and the systemic risk, and include open-ended testing of the model to improve the understanding of the systemic risk, with a view to identifying unexpected behaviours, capability boundaries, or emergent properties. Examples of model evaluation methods are: Q&A sets, task-based evaluations, benchmarks, red-teaming and other methods of adversarial testing, human uplift studies, model organisms, simulations, and/or proxy evaluations for classified materials. Further, the design of the model evaluations will be informed by the model-independent information gathered pursuant to Measure 3.1.

**Appendix 3.1 Rigorous model evaluations**

Signatories will ensure that the model evaluations are conducted with high scientific and technical rigour, ensuring:

(1) internal validity;

(2) external validity; and

(3) reproducibility.

**Appendix 3.2 Model elicitation**

Signatories will ensure that the model evaluations are conducted with at least a state-of-the-art level of model elicitation that elicits the model's capabilities, propensities, affordances, and/or effects, by using at least state-of-the-art techniques that:

(1) minimise the risk of under-elicitation; and

(2) minimise the risk of model deception during model evaluations (e.g. sandbagging);

such as by adapting test-time compute, rate limits, scaffolding, and tools, and conducting fine-tuning and prompt engineering.

For this, Signatories will at least:

(1) match the model elicitation capabilities of misuse actors relevant to the systemic risk scenario (pursuant to Measure 2.2); and

(2) match the expected use context (e.g. equivalent scaffolding and/or tool access) of the model, as informed by integrations into AI systems that are:

(a) planned or considered for the model; and/or

(b) currently used for similar models, if such integrations are known to the Signatory and the Signatory cannot exclude a similar use of their model.

### *Glossary*

External Validity: An aspect of high scientific and technical rigour (see definition below) that ensures model evaluations are suitably calibrated for results to be used as a proxy for model behaviour outside the evaluation environment. Demonstrating external validity will differ for different systemic risks and model evaluation methods, but may be shown by, e.g. documenting the model evaluation environment, the ways in which it diverges from the real-world context, and the diversity of the model evaluation environment.

**Appendix 3.4 Qualified model evaluation teams and adequate resources**

Model evaluation teams will be provided with:

(1) adequate access to the model to conduct the model evaluations pursuant to this Appendix 3.

## California Senate Bill 53 (2025-2026)

The following is an excerpt from California SB 53 (Senate Bill 53).

**22757.12.**

(a) A large frontier developer shall write, implement, comply with, and clearly and conspicuously publish on its internet website a frontier AI framework that applies to the large frontier developer's frontier models and describes how the large frontier developer approaches all of the following:

(1) Incorporating national standards, international standards, and industry-consensus best practices into its frontier AI framework.

(2) Defining and assessing thresholds used by the large frontier developer to identify and assess whether a frontier model has capabilities that could pose a catastrophic risk, which may include multiple-tiered thresholds.

…

(4) Reviewing assessments and adequacy of mitigations as part of the decision to deploy a frontier model or use it extensively internally.

(5) Using third parties to assess the potential for catastrophic risks and the effectiveness of mitigations of catastrophic risks.

## New York RAISE Act

The following is an excerpt from New York's RAISE Act (2025).

**A. 6453--B**

"Safety and security protocol" means documented technical and organizational protocols that:

(c) Describe in detail the testing procedure to evaluate if the frontier model poses an unreasonable risk of critical harm and whether the frontier model could be misused, be modified, be executed with increased computational resources, evade the control of its large developer or user, be combined with other software or be used to create another frontier model in a manner that would increase the risk of critical harm;

## Abbreviations

| | |
|---|---|
| AI | artificial intelligence |
| CBRN | chemical-biological-radiological-nuclear |
| CWM | closed-weight AI model |
| MFT | malicious fine-tuning |
| OWM | open-weight AI model |
| PE | proportional evaluation |
| RF | risk factor |
| RLHF | reinforcement learning from human feedback |

## About the Authors

**Patricia Paskov** is an Oxford Martin AI Governance Research Affiliate and an Adjunct Researcher at RAND. She conducts technical and policy research on the science of AI capability evaluations. Paskov holds a M.Res. and M.Sc. in Economics.

**Christopher Rodriguez** is a Biosecurity Research Scientist at RAND. He conducts technical and policy research on topics such as AI capability evaluations and biosecurity. Rodriguez holds a Ph.D. in Computational Biology.

**Sunishchal Dev** is an AI Evaluations Research Scientist at RAND. He conducts technical and policy research on such topics as AI capability evaluations and AI's implications on national security. Dev holds a B.A. in technology and innovation management.

**Stephen Casper** is a Fellow at the Harvard Berkman Klein Center and a final-year PhD student at MIT in Computer Science. He conducts technical and policy research on such topics as technical AI safeguards and governance. Casper holds a M.Sc. in Computer Science.